# Анализ на приложенията, подходящи за мобилно обучение на деца в предучилищна възраст

Александар Стоименовски, Радослава Кралева, Велин Кралев

*Катедра по информатика, Природо-математически факултет, Югозападен университет „Неофит Рилски"*

***Резюме***: *В тази статия са разгледани възможностите за прилагане на мобилно обучение в България, при деца в предучилищна възраст. Анализирани са най-използваните мобилни операционни системи. Представени са и са класифицирани някои от най-използваните съществуващи приложения, подходящи за обучение на малки деца. Направени са съответните изводи.*

***Ключови думи:*** *Мобилни приложения за деца в предучилищна възраст, Мобилно обучение*

## 1. ВЪВЕДЕНИЕ

Мобилните технологии са неделима част от съвременния начин на живот. Обикновено хората разполагат освен с телефон (най-често смартфон), още с таблет и компютър. Тези компютърни устройства са интересни и привлекателни за децата, защото работата с тях е лесна и интуитивна. Те ги използват за игри, за гледане на филми, слушане на музика, разговори с приятели и връзка със света около тях.

До преди няколко години единствените практики на преподаване, свързани с някакъв вид компютърно устройство, бяха електронното обучение (e-learning) и обучение през целия живот (u-learning). Към настоящия момент вече се говори и за друг по осъвременен начин на преподаване, а именно за мобилно обучение. Под този термин е залегнала идеята за използване на мобилни приложения, инсталирани на някакъв вид мобилно устройство, което се използва в задължителния обучителен процес на ученици.

Редица учени, като психолози, педагози и софтуерни разработчици започнаха изследвания свързани с използването на мобилните устройства, като средство за обучение на малки деца [1], [2], [3].

Мнението на родителите е важна част от процеса свързан с прилагането на този нов вид обучение. За да се отрази обективно тяхното отношение по този въпрос беше направено анкетно проучване, сред родители на деца на различна възраст и резултатите бяха отразени в [4]. Анализът на получените данни от това проучване е много удовлетворителен, тъй като се оказа, че повече от 90% от родителите подкрепят идеята за използване на мобилно обучение.

Ето защо може да се обобщи, че целта на настоящата статия е да се анализират наличните приложения за мобилни устройства в магазините на Google, Apple и Microsoft, които могат да бъдат използвани за мобилно обучение на деца в предучилищна възраст.

## 2. СЪЩЕСТВУВАЩИ ПРИЛОЖЕНИЯ, ПОДХОДЯЩИ ЗА МОБИЛНО ОБУЧЕНИЕ

Според [6] най-разпространените мобилни операционни системи в България са Android (71%) на Google Inc., iOS (54%) на Apple и Windows Phone (22%) на Microsoft. Сборът на процентите е по-голям от 100%, защото голяма част от хората използват повече от едно устройство, респективно повече от една операционна система.

Според статистика предложена в [7] голям дял от свалените приложения се пада на безплатните, като за 2015 година в световен мащаб, те са били малко над 167 милиарда, а платените – едва 12 милиарда и половина. Следователно може да се заключи, че компаниите, независимите разработчици и потребителите, залагат на безплатните приложения.

Ето защо в практиката с цел печалба, често се набляга на други похвати, като появата на нежелани реклами, или допълнителни модули, за които потребителят трябва да плати. Но всички те изискват непрекъсната връзка с интернет.

Най-голям дял от всички свалени приложения заемат игрите. Малък е относителния дял на софтуерните продукти, предназначени за обучение. А още по-малка част от тях могат да бъдат използвани безопасно от малки деца.

В този раздел ще бъдат изследвани именно онези приложения, налични в различните магазини на Google, Apple и Microsoft, които биха могли да бъдат използвани като средство за мобилно обучение на деца в предучилищна възраст. Всички разгледани приложения ще бъдат представени накратко и ще бъдат категоризирани.

Към днешна дата в магазина на Google Play има най-голямо разнообразие от приложения, които могат да бъдат използвани за обучението на деца в предучилищна възраст. Някои от най-сваляните са представени в Табл. 1. Голяма част от тези игри са безплатни за сваляне, но модулите предоставящи повече функции са платени. Основен проблем се оказва наличието на изскачащи реклами, които изискват непрекъсната връзка към интернет. Следователно не е осигурена безопасна среда за работа на децата. Освен това те покриват само малка част от изискванията на Държавните образователни изисквания (ДОИ) [5] и ако се пристъпи към едно токова мобилно обучение, ще бъде необходимо свалянето и инсталирането на много приложения, така че да се покрие целият изучаван материал. Някои от тези приложения са на български език с дружелюбен и интуитивен интерфейс.

*Таблица 1:* Някои от най-сваляните приложения в Google Play Store.

| Приложение | Цена | Поддържани езици | Операционна система | Умения | Възраст |
|---|---|---|---|---|---|
| Букви, цифри, цветове | Безплатно | Български | Android | Българските букви, цифри и цветове. | 5 – 7 год. |
| БГ Цифри | Безплатно | Български | Android | Цифри и броене до 20. | 2 – 5 год. |
| БГ Буквар | Безплатно | Български | Android | Български буквар с картинки, звуци и игри. | 2 – 6 год. |
| Kids: Preschool games | Безплатно | Английски, португалски, испански | Android | Букви, цифри, животните и техните звуци. | 5 – 7 год. |
| Animals memory games for kids | Безплатно | Английски | Android | Игра за развитие на памет. | 0 – 5 год. |
| Math games: Numbers for kids | Безплатно | Немски, испански, английски, японски, китайски, френски. | Android | Цифри, събиране и изваждане. | 6 – 8 год. |
| King of Math Junior | Безплатно | Американски английски | Android | Цифри, събиране и изваждане. | 6 – 8 год. |
| Kids: Coloring game | Безплатно | Английски | Android | Рисуване и оцветяване. | 1 – 4 год. |
| Letter School – learn write abc | 4.16 $ | Английски, немски, френски, испански | Android | Букви, цифри и азбука. | 5 – 7 год. |

Към момента на писане на статията в App Store на Apple бяха открити различни приложения, като някои от тях бяха аналогични на тези в Google Play (Табл. 2). Не бяха открити такива на български език за развиване на математическите познания на децата. Следователно не са покрити всички изисквания според ДОИ. Освен това приложенията на български език са много малко, не повече от 10 на брой и препокриват своите функционални възможности. Платените приложения тук са освободени от всякакви реклами и предоставят една по-спокойна среда за работа. Приложенията са с интуитивен и приятен дизайн.

*Таблица 2:* Някои от най-сваляните приложения в App Store.

| Приложение | Цена | Поддържани езици | Операционна система | Умения | Възраст |
|---|---|---|---|---|---|
| Буквите | Платено | Български | iOS 3.0 и по-нова | Българската азбука, начин на произнасяне и правопис на думи. | 6 – 8 год. |
| Българските букви цифри | Безплатно | Български | iOS 3.0 и по-нова | Букви, цифри и цветове. | 0 – 5 год. |

| Приложение | Цена | Поддържани езици | Операционна система | Умения | Възраст |
|---|---|---|---|---|---|
| Букви с витамини | £ 1.49 | Български | iOS 3.0 и по-нова | Произношение и начин на писане на българските букви. | 0 – 8 год. |
| Епълки | £ 2.29 | Български | iOS 5.0 и по-нова, iPhone, iPad | Пъзели за определяне на емоционалното състояние на децата. | 0 – 8 год. |
| Preschool! All in one | Безплатно демо | Американски английски | iOS 5.11 и по-нова | Букви, цифри, фигури. | 0 – 5 год. |
| Splash Math | Безплатно демо | Американски английски | iOS 7 и по-нова | Колекция от математически задачи, покриващи стандартната учебна програма, геометрия, умножение, деление. | 6 – 8 год. |
| Splash Math – 3rd Grade | Безплатно демо | Американски английски | iOS 7.0 и по-нова | Колекция от математически задачи. | 7 – 9 год. |

Както и при приложенията за Android, така и тук липсва едно цялостно приложение, което да обединява всички изисквания на ДОИ. Това важи за всички приложения с интерфейс на български език, които удовлетворяват само едно отделно направление.

В магазина Windows Store на Microsoft бяха разгледани няколко от най-често сваляните приложения, които биха могли да се използват като средство за мобилно обучение на малки деца и деца в предучилищна възраст. Някои от приложенията са представени в Табл. 3.

За разлика от останалите магазини тук не бе намерено нито едно приложение, което да поддържа интерфейс на български език. Всички приложения са с дружелюбен и интуитивен интерфейс.

Единственото приложение, което до голяма степен удовлетворява ДОИ е "Kids Play & Learn", но минусът се състои в липсата на поддръжка на български език. Освен това при всички приложения се използва връзка към интернет, което отново е съществен недостатък от гледна точка на за сигурността на децата.

*Таблица 3:* Някои от най-сваляните приложения в Windows Store

| Приложение | Цена | Поддържани езици | Операционна система | Умения | Възраст |
|---|---|---|---|---|---|
| Kids Play & Learn | Безплатно | Американски английски, немски, испански, италиански, полски, португалски, френски, руски, китайски и турски | Windows Phone 8 и по-нова | Изучаване на цветове, обекти, арабските и римските числа, броене, разпознаване на животни, редене на пъзели и други. | 2 – 10 год. |
| Kids Play Math | Безплатно | Американски английски | Windows Phone 8 и по-нова | Изучаване на числата, броене, събиране, изваждане, умножение, деление, игри от тип памет. | До 12 години |
| Games for Kids | Безплатно демо, $1.29 | Американски английски, немски, испански, италиански, полски, португалски, френски, руски, китайски и турски | Windows Phone 8 и по-нова | Разпознаване на обекти, цветове, числа и букви. | До 6 години |

| Приложение | Цена | Поддържани езици | Операционна система | Умения | Възраст |
|---|---|---|---|---|---|
| Kids Learning Word Games | Безплатно демо, $1.99 | Американски английски, немски, испански, полски, италиански, руски, португалски, френски, китайски и турски | Windows Phone 8.1 и по-нова | Основни познания по английски език, обогатяване на речника, класифициране, симетрия, памет. | За всички възрасти |
| Kindergarten Kids Learning | Безплатно демо, $1.99 | Американски английски, немски, испански, полски, италиански, турски, португалски, руски, френски и китайски. | Windows Phone 8 и по-нова | Тестове по разпознаване на букви, цифри, рими. | 2 – 5 год. |
| Kids Preschool Learning Games | Безплатно демо, модули $1.99, пълна версия $3.49 | Американски английски, немски, испански, италиански, полски, португалски, френски, руски, китайски и турски | Windows Phone 8 и по-нова | Основни математически познания, обогатяване на речника, класифициране, симетрия, намиране на съответствия. | 3 – 6 год. |
| Play School | Безплатно | Американски английски | Windows Phone 8 и по-нова | Азбука, числата от 1 до 20, прости фигури, цветове и животни. | 0 – 6 год. |

Изследването, което беше представено в настоящият раздел показа, че има голямото разнообразие от различни и не чак толкова различни приложения в магазините на Google, Apple и Microsoft. Повечето от тях притежават красив и интуитивен интерфейс. При по-голямата част от разгледаните приложения се изисква постоянна връзка към интернет. В голяма част от безплатните приложения са налични реклами, които пречат на нормалния начин на работа с приложението. Това говори за липса на надеждност и сигурност при тези приложения.

Някои от приложенията имат многоезикова поддръжка, но при повечето липсва български език, като за мобилната операционна система Windows Phone изцяло липсва такива. Освен това разгледаните приложения удовлетворяват учебния материал посочен в ДОИ, само в много малка част, като повечето не могат да покрият дори едно направление.

Ето защо за прилагане на съвременно мобилно обучение в предучилищният етап към този момент със съществуващите мобилни приложения в магазините на Google, Apple и Microsoft, е все още рано да се говори. За постигането на тази цел е необходимо да бъде разработено подходящо приложение, което да обхваща целият материал на ДОИ за деца в предучилищна възраст и осигурява безопасна среда на работа.

## 3. ЗАКЛЮЧЕНИЕ

Като резултат от изследването представено в тази статия, може да се заключи, че проблемът свързан с качественото разработване на програмни продукти за мобилни устройства, ориентирани към обучителния процес или за неговото подпомагане, все още съществува. В повечето случаи разработчиците целят бърза печалба, разчитайки на рекламодатели или на платени модули.

В тази статия бяха анализирани най-използваните мобилни операционни системи в България. Бяха класифицирани и анализирани някои от най-използваните мобилни приложения, подходящи за обучение на деца, в следствие на което бяха направени конкретни изводи.

## 4. ЛИТЕРАТУРА